\documentclass[aps,prl,reprint,groupedaddress,nofootinbib]{revtex4-1}
\usepackage{amsmath,amssymb}
\usepackage{graphicx}
\usepackage{bbm}
\usepackage{xcolor,footnote} 
\begin{document}

\title{Finite density condensation and scattering data - a study in $\phi^4$ lattice field theory}

\author{Christof Gattringer, Mario Giuliani, Oliver Orasch}

\affiliation{Universit\"{a}t Graz, Institut f\"{u}r Physik, Universit\"{a}tsplatz 5, 8010 Graz, Austria}

\date{26.2.2018}

\begin{abstract}
We study the quantum field theory of a charged $\phi^4$ field in lattice regularization 
at finite density and low temperature in 2 and 4 dimensions with the goal of analyzing the connection of 
condensation phenomena to scattering data in a non-perturbative way. The sign problem 
of the theory at non-zero chemical potential $\mu$ is overcome by using a worldline representation for
the Monte Carlo simulation. At low temperature we study the particle number as a function of $\mu$ 
and observe the steps for $\mbox{1-,}$ 2- and 3-particle condensation. We determine the corresponding 
critical values $\mu_n^{crit}, \, n = 1,2,3$ and analyze their dependence on the spatial extent $L$ of the 
lattice. Linear combinations of the $\mu_n^{crit}$ give the interaction energies in the 2- and 3-particle sectors 
and their dependence on $L$ is related to scattering data by L\"uscher's formula and its generalizations 
to three particles. For 2-$d$ we determine the scattering phase shift and for 4-$d$ the scattering length. 
We cross-check our results with a determination of the mass and the 2- and 3-particle energies from conventional 
2-, 4-, and 6-point correlators at zero chemical potential. The letter demonstrates that the physics of 
condensation at finite density and low temperature is closely related to scattering data of a quantum field theory.
\end{abstract}

\maketitle

\section{Introduction}
It is well known that phenomena in low energy physics can be described in terms of a few low energy parameters, which 
in the context of this letter was, e.g., discussed in the seminal paper \cite{huang_yang}. As a particular 
instance of this relation the condensation of particles at low temperature and non-zero chemical potential can be 
related to scattering data of the corresponding quantum field theory. More specifically one may show 
\cite{Bruckmann} that the critical 
values of the chemical potential where one observes condensation steps at low temperature and small volume are related 
to the finite volume many particle energies, which in turn are related to the 
scattering length \cite{huang_yang}.

For studying low energy properties non-perturbative methods need to be applied, e.g., lattice simulations. 
However, for many finite density lattice field theories the action is complex. The 
Boltzmann factor has a complex phase and cannot be used as a probability in a lattice Monte Carlo study. 
This ''sign problem'' has 
recently been overcome for several theories by exactly mapping them to a worldline representation with only real
and positive weights (see \cite{Endres,Weisz,phi4_1,phi4_2,preliminary} for the model considered here). 

In this letter we study the complex $\phi^4$ field at finite chemical potential using a worldline representation. We 
compute the particle number as a function of the chemical potential at low temperature and determine the condensation 
steps for the 1-, 2- and 3- particle sectors. Analyzing their volume dependence we show non-perturbatively 
that the condensation steps are indeed related to the scattering data of the theory.

\section{Worldline representation}
The dynamical 
degrees of freedom of the charged $\phi^4$ 
field are the complex valued fields $\phi_x$ assigned 
to the sites $x$ of a $d$-dimensional lattice with periodic
boundary conditions. Here we consider $d = 2$ and $d=4$, 
i.e., we work on lattices with volumes 
$V = N_s \times N_t$ and $V = N_s^3 \times N_t$, where $N_s$ 
is the spatial extent in lattice units. $N_t$ is the extent in Euclidean 
time direction (the $d$ direction), which also equals the inverse 
temperature $\beta$ in lattice units. The grand canonical partition sum
is given by the path integral $Z   = \int D[\phi] e^{-S[\phi]}$ with the 
product measure $\int \! D[\phi] = \prod_x \int_\mathbb{C} d \phi_x/2\pi$.
The lattice action is given by
\begin{eqnarray}
S[\phi]  & = &  
\sum_{x \in \Lambda} \bigg( \eta \, |\phi _{x}|^2 \; + \; \lambda \, |\phi _{x}|^4  
\label{action_conventional} \\
& & \qquad - \sum_{\nu = 1}^{d} 
\left[ e^{\,\mu \delta_{\nu,d}}\phi _{x}^{\ast} \phi _{x+\hat{\nu}} + 
e^{\,-\mu \delta_{\nu,d}}\phi _{x+\hat{\nu}}^{\ast} \phi _{x} \right]\bigg) \; .
\nonumber 
\end{eqnarray}
The bare mass $m_b$ enters via the parameter $\eta \equiv 2d + m^{2}_b$, 
$\lambda$ denotes the coupling of the quartic self-interaction and $\mu$ 
is the chemical potential. 

For nonzero $\mu$ the action (\ref{action_conventional}) is complex 
and the model has a sign problem in the conventional formulation.
The sign problem can be solved by exactly mapping the theory to a 
worldline representation where also at finite $\mu$
all weights are real and positive, such that a simulation is possible 
directly in terms of the worldlines \cite{Endres,Weisz,phi4_1,phi4_2,preliminary}. 
In the worldline form the partition sum is given by
\begin{equation}
Z   \; = \; \sum_{\{k\}}  \left[ \prod_{x} \delta \left(\vec{\nabla} \cdot \vec{k}_{x} \right) \right] \, 
e^{\, \mu \, \beta \, \omega[k]} \; B[k] \; .
\label{worldlineZ}
\end{equation}
The sum is over all configurations of the worldline variables $k_{x,\nu} \in \mathbb{Z}$ 
assigned to the links of the lattice. The worldline variables 
$k_{x,\nu}$ obey a zero divergence constraint, which we write as a product over all lattice sites 
and at each site $x$ a Kronecker 
delta $\delta(j) \equiv \delta_{j,0}$ enforces vanishing divergence
$\vec{\nabla} \cdot \vec{k}_{x}  \equiv   \sum_{\nu}(k_{x,\nu} - k_{x-\hat{\nu},\nu}) = 0$.
Consequently the worldline variables $k_{x,\nu}$ must form closed loops of conserved flux.

The chemical potential $\mu$ couples to the temporal winding number $\omega[k]$ of the conserved flux: 
By comparing the $\mu$-dependent 
term in (\ref{worldlineZ}) to the standard form $e^{\mu \beta N}$ for the $\mu$-dependence of the 
grand canonical partition sum, we conclude that the net-particle number $N$ is given by the 
temporal net winding number $\omega[k]$ of the worldlines.  Finally each configuration comes with a weight factor 
\begin{eqnarray}
&& 
B[k] \; \equiv \; \sum_{\{a\}} \, \prod_{x,\nu}\frac{1}{(a_{x,\nu}+|k_{x,\nu}|)! \, a_{x,\nu}!} \,  \prod_{x} I(s_x) \; ,
\label{weights} \\
&& 
\mbox{with} \; \; I(s_x) \; = \; \int_{0}^{\infty} \!\! d r \; r^{\, s_x + 1} \, e^{\, -\eta \, r^2 \, - \, \lambda \, r^4} \; .  
\nonumber
\end{eqnarray}
The weight factor is itself a sum over configurations $\sum_{\{a\}}$ of
auxiliary link variables $a_{x,\nu} \in \mathbb{N}_0$.
The integrals $I(s_x)$ come from integrating out the radial degrees of freedom of the original field variables at site $x$.  
The argument $s_x$  is a non-negative integer combination of the auxiliary 
variables and the moduli of all $k$-fluxes that run
through $x$, defined as $s_x \; = \; \sum_{\nu}\Big[|k_{x,\nu}| +  |k_{x-\hat{\nu}}| + 2(a_{x,\nu} +  a_{x-\hat{\nu}})\Big] $. 
For the numerical simulation the integrals $I(s_x)$ are pre-calculated and 
stored for sufficiently many values of the arguments 
$s_x \in \mathbb{N}_0$. In a Monte Carlo simulation of the worldline form the  
variables $k_{x,\nu}$ and the auxiliary variables $a_{x,\nu}$ need to be updated, 
such that the $k_{x,\nu}$ obey all constraints. We apply a strategy based on the worm algorithm 
\cite{worm} with the details discussed in \cite{algorithm}.

For the 4-$d$ simulations we use lattices with $N_t = 320$ and 640, and $N_s$ between 3 and 10 at coupling values of 
$\eta = 7.44$ and $\lambda = 1.0$. The statistics is $10^5$ to $2 \times 10^5$ configurations. 
For 2-$d$ we use $N_t = 400$ and 
$N_s$ between 2 and 16 with $\eta = 2.6$, $\lambda = 1.0$ and a statistics of $4 \times 10^5$. 

We conclude this section with an estimate of discretization effects for the 4-$d$ case. As we will see below, 
at the couplings we use the infinite volume mass 
$m_\infty$ of the lowest excitation is $m_\infty \sim 0.168/a$, where $a$ is the lattice constant. 
For the momentum cutoff $p_{c} = \pi/a$ we thus find $p_{cut}/m_\infty = \pi/0.168 \sim 18.9$. 
The largest energy we use in our study is the 3-particle energy $W_3$ for $N_s = 4$, which at $W_3(4) \sim 0.93/a$ is
less than 6 times $m_\infty$. We expect that with $p_c/W_3(4) = \pi/0.93 \sim 3.4$ cutoff effects are small. A 
similarly good ratio holds for our 2-$d$ study.

\section{Condensation thresholds and multi-particle energies}

Using the worldline representation summarized in the previous section we now analyze the particle number $N$ 
as a function of the chemical potential $\mu$. We have already discussed that in the 
worldline representation the particle number $N$ is represented by the temporal 
winding number $\omega[k]$ of the worldlines, and thus we study the expectation value 
$\langle N \rangle \equiv \langle \omega[k] \rangle$,
where the vacuum expectation value on the rhs.\ of this equation is evaluated in the worldline representation. 

\begin{figure}[t!]
\begin{center}
\includegraphics[height=10cm]{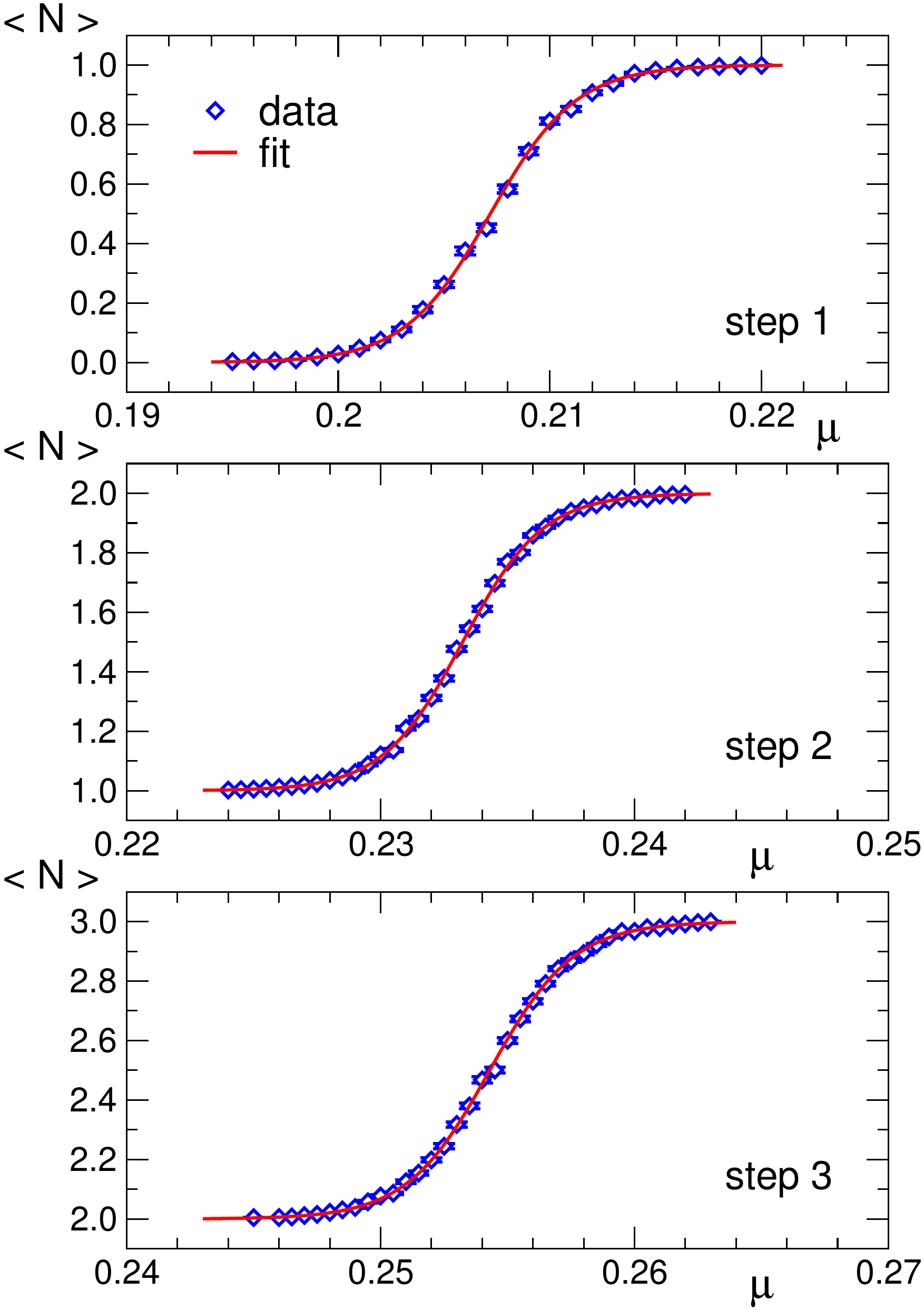}
\end{center}
\vspace*{-6mm}
\caption{The expectation value of the particle number $\langle N \rangle$ as a function of $\mu$ for the 
4-$d$ case with $N_s = 6$. We show the regions of 1- (top plot) 2- (middle) and 3-particle condensation (bottom). 
The Monte Carlo data (symbols) are fit with the logistic function (full curve).}
\label{steps}
\end{figure}

In Fig.~\ref{steps} we show the expectation value of the particle number 
$\langle N \rangle$ as a function of the chemical potential $\mu$ for the 
4-$d$ case with $N_s = 6$. We split the figure into 3 plots, choosing the ranges of $\mu$ such that we see the 
condensation thresholds for 1-, 2- and 3-particle condensation, i.e., the values $\mu_n^{crit}, n = 1,2,3$ 
where $\langle N \rangle$ quickly 
climbs from $\langle N \rangle = n-1$ to $\langle N \rangle = n$. 
One observes that $\langle N \rangle$ has shoulders at the
integers $0,1,2,3$ and shows a rapid increase in between. 
In the zero temperature limit, i.e., for $N_t \rightarrow \infty$ this turns into steps,
which here at finite $N_t$ are rounded due to temperature effects. 

Near the transition from $\langle N \rangle = n\!-\!1$ to $\langle N \rangle = n$ we fit the data with a 
logistic function shifted by a constant, $\langle N \rangle \!=\! [1\!\,+\!\, \exp(-a_{n}[\mu\!-\!\mu_n^{crit}])]^{-1}\!+\!n\!-\!1$. 
The plots show that this 2-parameter ($a_n$ and $\mu_n^{crit}$) fit function describes the data very well and 
allows us to determine the critical values $\mu_n^{crit}, n = 1,2,3$ of the chemical potential. As a 
cross-check we determined the critical values $\mu_n^{crit}$ also from the peaks of the particle number 
susceptibility and found excellent agreement of the two determinations.  

In \cite{Bruckmann} it was pointed out, that the critical values $\mu_n^{crit}, n = 1,2,3$ at low 
temperature are related to the 1-, 2- and 3-particle energies at finite volume, i.e.,
\begin{eqnarray}
\mu_1^{crit} & = & m \; ,
\nonumber \\
\mu_1^{crit} + \mu_2^{crit} & = & W_2 \; ,
\nonumber \\
\mu_1^{crit} + \mu_2^{crit} + \mu_3^{crit} & = & W_3 \; ,
\label{mucombinations}
\end{eqnarray}
where $m$ is the renormalized physical mass, $W_2$ the 2-particle energy, and $W_3$ the 3-particle energy. 

\begin{figure}[t!]
\begin{center}
\includegraphics[height=7.5cm]{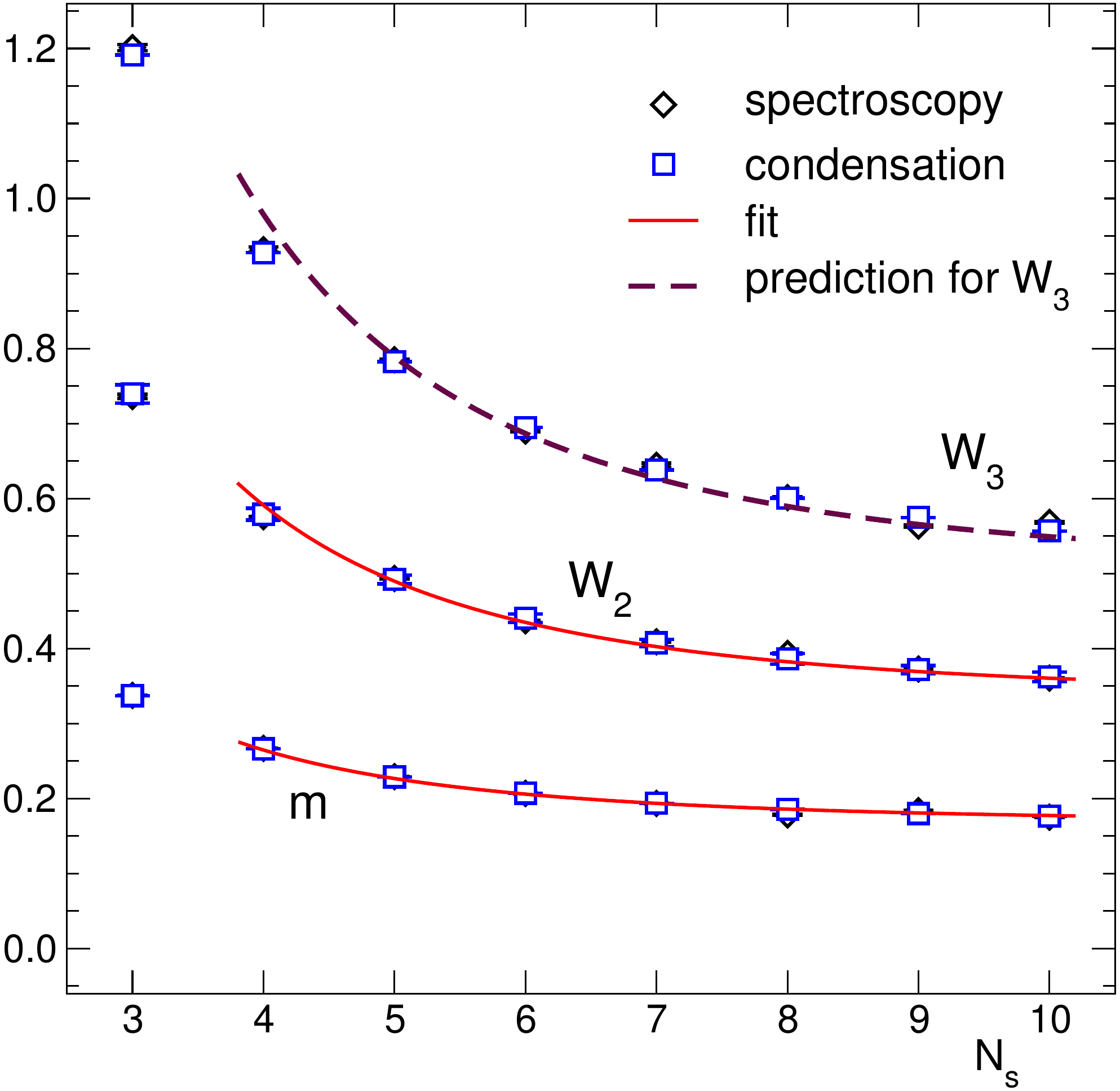}
\end{center}
\vspace*{-6mm}
\caption{The physical mass $m$ and the 2- and 3-particle energies $W_2$ and $W_3$ for the 4-$d$ case 
as a function of the spatial extent $N_s$. We compare the results from the $n$-particle energy thresholds (squares) to 
the results from a direct determination with correlators (diamonds). The fits to $m$ and $W_2$ are shown as solid 
red curve, and the prediction for $W_3$ based on these fits is shown as a dashed purple curve.}
\label{EversusL}
\end{figure}

When changing the spatial extent $N_s$ of the 
lattice one observes a shift of the condensation steps $\mu_n^{crit}$ and thus
of $m$, $W_2$ and $W_3$. In Fig.\ \ref{EversusL} we show the 
$N_s$-dependence of these three quantities (squares), 
which in the next section we will use to determine scattering data. 

Before we come to the analysis of the finite volume effects and their 
connection to scattering data we present an important cross-check of
our worldline results. The mass $m$, as well as $W_2$ and $W_3$ can also
be computed from the exponential decay of Euclidean $2n$-point functions in the 
conventional representation. More specifically we consider the spatially Fourier transformed 
fields at zero momentum, $\widetilde{\phi}_t \; = \; (N_s)^{-3} \sum_{\vec{x}} \phi_{\vec{x},t}$, 
and compute the connected $2n$-point functions for $n = 1,2,3$:
\begin{equation}
\langle \big( \widetilde{\phi}_t \big)^n \; \big( \widetilde{\phi}_0^\star \big)^n \rangle_c \; \propto \; A \, e^{-t E_n} \; ,
\label{correlators}
\end{equation}
where $E_1 = m$ and $E_n = W_n$ for $n = 2,3$. From a fit to the correlators we determined the values
for $m$, $W_2$ and $W_3$, which in Fig.~\ref{EversusL} are shown as diamonds. As a cross-check we determined 
$m$, $W_2$ and $W_3$ also using a full correlation matrix and got values agreeing very well 
with those from the $2n$-point functions (\ref{correlators}), which indicates that contributions of
excited states are negligible. 

Fig.~\ref{EversusL} shows that the values for $m$, $W_2$ and $W_3$ as determined from the condensation steps 
agree very well with the values from the $2n$-point functions. This establishes that the condensation steps
observed for the particle number are indeed determined by the corresponding $n$-particle energies via the 
relations (\ref{mucombinations}). The same comparison was done also for the 2-$d$ case and again we found excellent 
agreement of the energies determined from the critical  $\mu_n^{crit}$ and those from the $2n$-point functions.

\section{Volume dependence and scattering}

Having shown that the low temperature condensation steps are indeed governed by the $n$-particle energies
we can now apply known finite volume relations to connect the condensation steps with scattering data. 

\begin{figure}[t!]
\begin{center}
\includegraphics[height=57mm,clip]{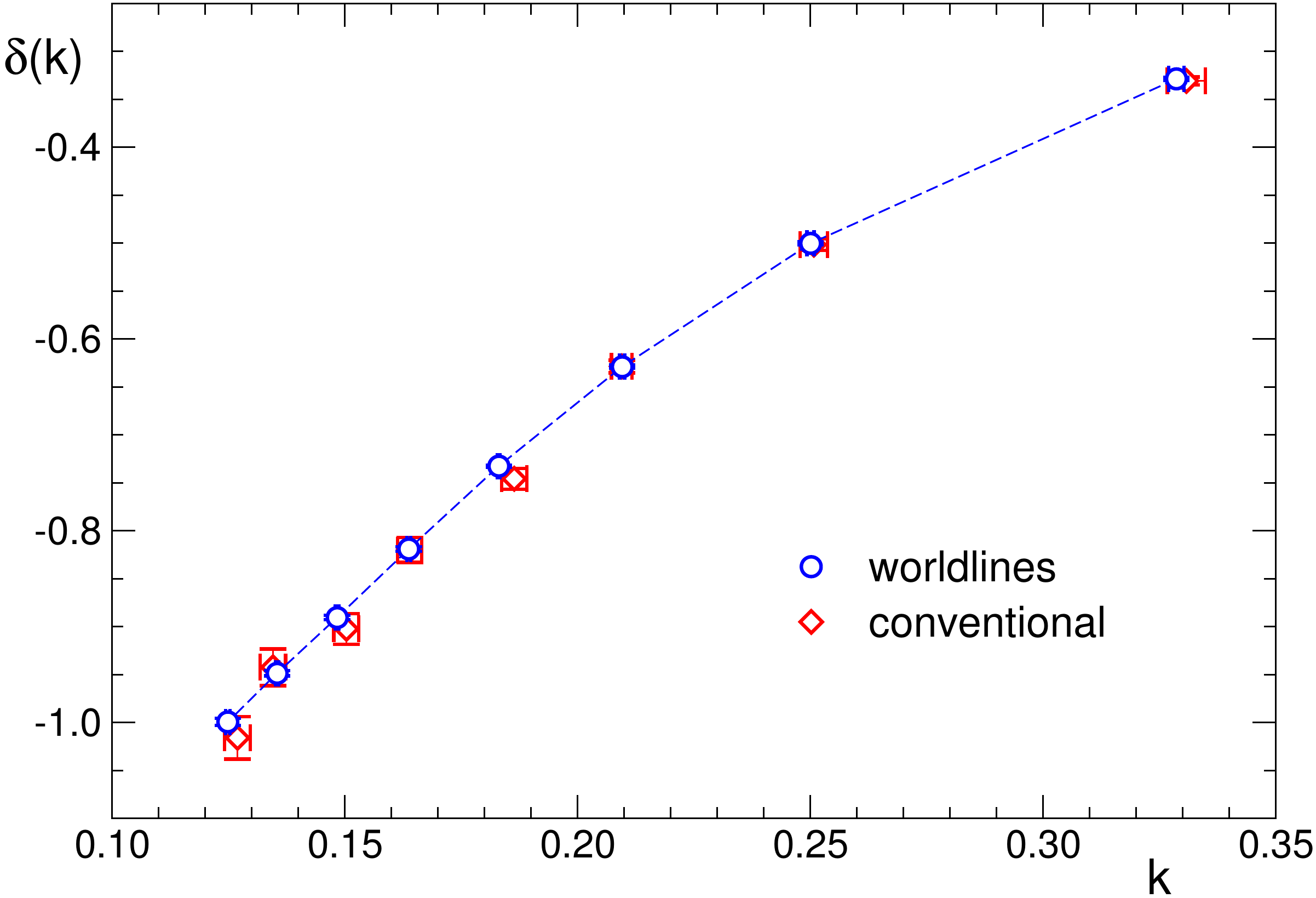}
\end{center}
\vspace*{-6mm}
\caption{Scattering phase shift $\delta(k)$ for 2-$d$ 
as a function of the momentum $k$. We compare the results from the worldline calculation to the data
from the conventional approach. 
}
\label{phaseshift2d}	
\end{figure}

We begin this analysis with the 2-$d$ case following \cite{Luscher_2d}.  There 
the 2-particle energy $W_2$ is related to the relative momentum $k$ of the particles via
$W_2  =  2\sqrt{ m^2 + k^2}$, and we can invert this equation to determine the momentum 
$k$ for each value of $W_2$. On our finite lattice with spatial extent $N_s$ the momentum $k$ is subject to the 
quantization condition $e^{2i\delta(k)} \; = \; e^{-i k N_s}$, where $\delta(k)$ is the phase shift for that momentum. 
Combining this relation with the relation between $W_2$ and $k$, we can extract the scattering phase shift 
$\delta(k)$ from $W_2$. Varying $N_s$ gives rise to different values of the relative momentum $k$ such that
$\delta(k)$ can be determined for a whole range of momenta. 

In Fig.~\ref{phaseshift2d} we plot the results for $\delta(k)$ as a function of $k$ and again compare the 
data determined from the $n$-particle energy thresholds ($m = \mu_1^{crit}$,  
$W_2 = \mu_1^{crit} + \mu_2^{crit}$) to the results obtained by determining $m$ from 2-point
functions and $W_2$ from 4-point functions. We find very good agreement of the two data sets and thus 
establish the relation of the condensation steps to the scattering phase shift for the 2-$d$ case.

For the 4-$d$ case we use the finite volume relations for $m$ \cite{kari}, the result \cite{huang_yang,Luscher_w2} for the 
2-particle energy $W_2$ (using the notation of \cite{sharpe}) and the results 
\cite{beane,sharpe,sharpe1,sharpe2,sharpe3} for the 3-particle energy $W_3$:
\begin{eqnarray}
\hspace*{-5mm}  m & = & m_\infty + \frac{A}{L^{\frac{3}{2}}} \, e^{ - L \, m_\infty } ,
\label{mL} \\
\hspace*{-5mm} W_2 & = & 2m  +  \frac{4\pi a}{m L^3} \! 
\Bigg[ 1 - \frac{a}{L} \frac{{\cal I}}{\pi} 
 + 
\bigg(\!\frac{a}{L}\! \bigg)^{\!\!2} \, \frac{ {\cal I}^{\,2} \!-\! {\cal J}}{\pi^2}  + 
{\cal O} \! \bigg(\!\frac{a}{L}\! \bigg)^{\!\!3}\Bigg]\! ,
 \label{W2} 
\end{eqnarray}
\begin{eqnarray} 
\hspace*{-5mm} W_3 & = & 3m +  \frac{12\pi a}{m L^3} \! 
\Bigg[ 1 - \frac{a}{L} \frac{{\cal I}}{\pi}  + 
\bigg(\!\frac{a}{L}\! \bigg)^{\!\!2} \, \frac{ {\cal I}^{\,2} \!+ \!{\cal J}}{\pi^2}  +  
{\cal O} \! \bigg(\!\frac{a}{L}\! \bigg)^{\!\!3}\Bigg]\! .
\label{W3}
\end{eqnarray}
Fit parameters are the infinite volume mass $m_\infty$, the amplitude $A$ and the scattering length $a$. 
The numerical constants ${\cal I}$ and ${\cal J}$ are given by ${\cal I} = -8.914, {\cal J} = 16.532$.

We identify $L \equiv N_s$ and fit the data for $m(N_s)$ as determined from 
$\mu_1^{crit}$ in the range between $N_s = 4$ and $N_s = 10$ with the functional form (\ref{mL}). This 
fit gives the amplitude parameter $A$ and the infinite volume mass $m_\infty$ in lattice units. Subsequently we 
fit $W_2(N_s)$ determined from $\mu_1^{crit} + \mu_2^{crit}$ with the functional form (\ref{W2}), again in the range 
between $N_s = 4$ and $N_s = 10$. We use $m(N_s)$ from the previous step, such that this second fit is a 
1-parameter fit that gives the scattering length $a$ in lattice units. From Fig.~\ref{EversusL} it is obvious that 
the fits for $m(N_s)$ and 
$W_2(N_s)$ describe the data very well and indeed the reduced $\chi^2$ is close to 1 for both fits. 

For $W_3(N_s)$ no additional free parameter is needed such that we simply can compare our data to the curve for $W_3$ 
that we obtain from (\ref{W3}) using the fit parameters of the previous fits as input. In Fig.~\ref{EversusL} 
we observe good agreement 
of this ''predicted'' $W_3$ with the data from the condensation steps. Only for the smallest $L$ $(\equiv N_s)$ we 
observe a deviation which shows that here higher order corrections in $a/L$ start to play a role. 
However, it has to be remarked that cleanly determining the higher order power law corrections in $W_2$ and $W_3$ or 
exponential terms of the form $e^{-L/R}$, where $R$ is the range for an interaction with an exponential tail, 
is a non-trivial numerical challenge. Such terms contribute significantly only for very small extent $L$ and for a 
serious determination one should resolve small $L$ with several data points, which in turn implies working with 
very fine lattices and very high statistics.  

For completeness we quote our results for the two physical parameters, i.e., the mass $m_\infty$ and the 
scattering length $a$. In lattice units we obtain $m_\infty = 0.168(1)$ and $a = - 0.078(7)$ and a value of 
$a \, m_\infty = - 0.013(1)$ for their dimensionless product. 

\section{Discussion and concluding remarks}

In this letter we determined the $n$-particle energy 
thresholds $\mu_n^{crit}, n = 1,2,3$ in a worldline simulation of the charged $\phi^4$ field at 
finite density. These thresholds correspond to the first three unit steps of the particle number expectation value 
$\langle N \rangle$  and emerge for small volumes and low temperatures. From the $\mu_n^{crit}$ we determined the
mass $m$, as well as the 2- and 3-particle energies $W_2$ and $W_3$ using (\ref{mucombinations}). We 
studied their dependence on the spatial extent $N_s$ and cross-checked the condensation results with those 
from $2n$-point functions. We found very good agreement of the data for all values of $N_s$ we analyzed, 
thus demonstrating that the condensation thresholds are indeed governed by the lowest $n$-particle energies, 
which correspond to the mass and 2- and 3-particle scattering states. This is the first time that this agreement 
is shown non-perturbatively in four dimensions, since the necessary worldline techniques for finite density lattice 
simulations became available only recently. 

Subsequently we analyzed the volume dependence of $W_2$ and $W_3$ in order to make contact to 
scattering data. For the 2-$d$ case such an analysis allows for a complete determination of the scattering phase 
$\delta(k)$ from $W_2$. In four dimensions we fit the $L$-dependence of $m$ and $W_2$ to determine
the infinite volume mass $m_\infty$ and the scattering length $a$. Inserting these parameters in the functional
form for $W_3$ determines the 3-particle energy up to $1/L^6$ corrections. We compared this ''prediction'' with 
our results for $W_3$ from the  condensation data and found very convincing agreement. This demonstrates that 
the finite density condensation steps at low temperature are indeed governed by the scattering data of the 
theory. 

We stress at this point that the scenario for the condensation which was exploited in our analysis requires a repulsive interaction. In case of an attractive interaction one expects condensation thresholds at the masses of the bound states,
and a nice example of this scenario is documented in \cite{G2QCD} for the model of QCD with the exceptional gauge group 
$G_2$, which is free of sign problems.

It is well known that in four dimensions $\phi^4$ theory is trivial. The renormalized coupling vanishes when 
approaching the continuum limit and one may expect that also the scattering length vanishes. Consequently 
our study in $\phi^4$ theory deals with an effective theory which, however, has a large scaling region where 
it essentially behaves like a continuum theory at low energies, 
as has been demonstrated in \cite{triviality}. In 2-$d$, on the other hand, 
one could try to run towards the Ising fixed point to construct a non-trivial continuum limit, where one expects a constant 
phase shift of $\delta(k) = - \pi/2$ \cite{ising_delta}.

Having established the interesting connection between scattering data and thermodynamical properties at low
temperature in a simple scalar theory it is of course interesting to ask if this can also been seen in more rich theories. 
The key issue is to be able to simulate the theory at finite density, which often is spoiled by the sign problem. However,
two interesting cases can be addressed immediately, namely QCD with only two colors and QCD with an isospin
chemical potential. In both cases the sign problem is absent, but the former case clearly is of more academic interest. 
However, using an isospin chemical potential one can condense pions in full QCD and study the relation of the second 
and third condensation thresholds to scattering data. This is a highly interesting connection that could be studied 
along the lines presented in this letter.

\vskip3mm
\begin{acknowledgments}
{\bf Acknowledgements:} 
We thank F.\ Bruckmann, T.\ Kloiber, A.\ Maas, C.B.\ Lang, C.\ Marchis, S.\ Sharpe, T.\ Sulejmanpasic, and U.\ Wolff 
for discussions. This work is supported by the Austrian Science Fund FWF, grant 
I 2886-N27 and the FWF DK W 1203, ''Hadrons in Vacuum, Nuclei and Stars". 	
\end{acknowledgments}


\end{document}